# Identification of formation of amorphous Si phase in SiO$_x$N$_y$ films produced by plasma enhanced chemical vapor deposition

M. V. Voitovych[a], A. Sarikov[a, b, c, *],  V. O. Yukhymchuk[a], V. V. Voitovych[d], M. O. Semenenko[a, c, d]

[a] V. Lashkaryov Institute of Semiconductor Physics, National Academy of Sciences of Ukraine, 41 Nauky Avenue, 03028 Kyiv, Ukraine

[b] Educational Scientific Institute of High Technologies, Taras Shevchenko National University of Kyiv, 4-g Hlushkova Avenue, 03022 Kyiv, Ukraine

[c] National Technical University of Ukraine "Igor Sikorsky Kyiv Polytechnic Institute", 37 Beresteiskyi Avenue, 03056 Kyiv, Ukraine

[d] Institute of Physics, National Academy of Sciences of Ukraine, 46 Nauky Avenue, 03028 Kyiv, Ukraine

*Corresponding author's email: sarikov@isp.kiev.ua

**Abstract**

Peculiarities of formation of inclusions of amorphous Si (a-Si) phase in Si-rich Si oxynitride films grown by plasma-enhanced chemical vapor deposition (PECVD) are studied by combined Raman scattering and infrared (IR) absorption spectroscopy. The Raman scattering results identify presence of a-Si phase in the studied films at the relative Si content exceeding a threshold value of about 0.4. The a-Si amount correlates with the concentration of hydrogen in the films, the presence of which is detected by characteristic IR absorption bands corresponding to Si–H bending (~660 cm$^{-1}$) and stretching (a composite band in the range of ~1900-2400 cm$^{-1}$) vibrations. The method of deconvolution of IR absorbance spectra in the range of ~600 to 1300 cm$^{-1}$ developed earlier is used to reliably separate the IR band at ~660 cm$^{-1}$. This band is identified to origin from the amorphous Si phase within the studied Si oxynitride films. This makes it possible to propose IR spectroscopy with analysis of the low-wavenumber part of the spectra as an efficient method of identifying phase composition of Si-rich Si oxynitride films. The obtained results contribute to understanding of the regularities of formation of phase compositions of PECVD grown Si oxynitride films and are useful for controlling the films properties for practical applications.

# 1. Introduction

Si oxynitride films (SiO$_x$N$_y$, $0 \leq x \leq 2$, $0 \leq y \leq 4/3$) have a great significance for fabricating modern microelectronic and optoelectronic devices [1, 2]. As the structure of such films defines their properties, their study attracted great attention in recent decades [3, 4]. Naturally, the main focus was to study the behavior of atoms of the main film forming elements, namely silicon, oxygen and nitrogen. It may be considered well-established today that structural arrangement of Si, O and N atoms in the SiO$_x$N$_y$ films depends on the film composition. In the films with less than ~50% of Si, the chemical composition is dominated by Si suboxide bonding (set of silicon-oxygen tetrahedra obeying the random bonding statistics) with N present as a significant impurity [5, 6]. Considering the N atoms entering into the network to substitute O atoms maintaining the same structure as that of a stoichiometric SiO$_2$ material, the proper chemical formula for such oxynitride materials should be SiO$_{2-y}$N$_y$ [7].

Increase of the relative Si concentration principally changes the SiO$_x$N$_y$ film morphology and structure. In [8], Si-rich films were considered as composites, the microstructure of which may be represented as a mixture of silicon-oxygen tetrahedra (Si–O$_a$Si$_{4-a}$, $a = 0…4$) and silicon-nitrogen pyramids (Si$_3$N) in the framework of the mixture model. Along with formation of such phase composition, existence of variable-size amorphous Si (a-Si) clusters embedded in a dielectric matrix in as-deposited Si-rich SiO$_x$N$_y$ films was inferred based on photoluminescence (PL) and Raman spectroscopy studies [9, 10]. In particular, these clusters were responsible for room-temperature PL in the range of 1.5-2.0 eV. Manifestation of two luminescence peaks [4] was attributed to two different average cluster sizes. The PL origin from the amorphous Si inclusions in this case was supported by the presence of the features of elemental Si in the X-ray photoelectron spectra of the samples with $x < 0.96$ and by a lack of crystalline features in the glancing angle X-ray diffraction patterns.

Infrared (IR) spectroscopy with mathematical deconvolution of the IR absorption band is successfully used for study of the microstructure of Si oxide films with excess Si content [11, 12]. In our previous publication [13], the characteristics of the elementary bands corresponding to Si–N and Si–H bond vibrations in the spectral range of ~600 to 1300 cm$^{-1}$ have been determined that enable reliable deconvolution of the IR absorbance spectra with separation of the Si–O, Si–N, and Si–H components. It is known that hydrogen in Si oxynitride films is contained in high (of the order of $10^{22}$ cm$^{-3}$ [7, 14]) concentrations and is responsible for quite noticeable absorption of IR radiation in well-known spectral regions. On the other hand, the Si–H bond vibration frequency is very sensitive to the local structural arrangement, which is evidenced by the results of the studies of such materials as Si oxides and amorphous Si [15-17]. This opens a way to investigate IR absorption spectroscopy as a method for identifying not only Si oxynitride, oxide and nitride phases but also amorphous Si phase in Si oxynitride films. In this work, we use combined Raman scattering and Fourier transform IR (FTIR) spectroscopy with analysis of various Si–H vibration bands to investigate the peculiarities of the formation of a-Si phase in the Si-rich Si oxynitride films obtained by plasma-enhanced chemical vapor deposition (PECVD).

## 2. Experimental

In our experiments, 300 ± 5 nm thick $SiO_xN_y$ films with different compositions (stoichiometry indices $x$ and $y$) obtained by PECVD were used. The values of the $N_2O/SiH_4$ flow ratio during the film deposition and the respective film stoichiometries are presented in Table 1. Double-side polished boron-doped CZ Si wafers with the thickness of 300 μm and 1 mm thick sapphire plates were used as substrates for infrared (IR) and Raman spectroscopy investigations, respectively. More details about the film deposition procedure and determination of the film composition can be found in our previous publication [13].

**Table 1.** Deposition conditions and composition of $SiO_xN_y$ films

| Sample | $N_2O/SiH_4$ flow ratio | Composition | | |
|---|---|---|---|---|
| | | $x$ | $y$ | Relative Si content |
| #1 | 9 | 1.95 | 0.01 | 0.34 |
| #2 | 3 | 1.3 | 0.28 | 0.39 |
| #3 | 1.5 | 1.09 | 0.32 | 0.41 |
| #4 | 0.6 | 0.85 | 0.27 | 0.48 |
| #5 | 0.4 | 0.59 | 0.16 | 0.57 |
| #6 | 0.3 | 0.42 | 0.14 | 0.64 |
| #7 | 0.2 | 0.37 | 0.11 | 0.68 |
| #8 | 0.1 | 0.23 | 0.07 | 0.77 |
| #9 | 0.06 | 0.18 | 0.05 | 0.80 |

Raman spectra of the studied films were excited by the emission of a CNI Model PSU-H-FDA solid-state laser with λ = 457 nm at room temperature and recorded using an MDR-23 spectrometer equipped with an Andor iDus 401A CCD detector (UK). In order to prevent the film structure from being affected during the measurements, the excitation radiation power density was fairly small, not exceeding $10^3$ W/cm². For correct comparison of the bands in different Raman spectra, the latter were normalized to the intensity of the band corresponding to amorphous Si (~480 cm$^{-1}$).

FTIR transmittance spectra in the range of 400−4000 cm$^{-1}$ were measured at room temperature using a PerkinElmer BX-II spectrometer. The measurement resolution was 2 cm$^{-1}$, the number of scans was 100, and the measurement accuracy was ~0.5%. A Si substrate was used as a reference sample for all the measurements.

IR absorbance spectra were recalculated from the transmittance spectra using the Beer–Bouguer–Lambert law. In this work, the bands related to Si–H stretching (a wide composite band at around 2200 cm$^{-1}$) and bending (peaked near 660 cm$^{-1}$) vibrations were analyzed. To separate the contributions from different Si–H containing complexes, the IR absorption bands corresponding to the Si–H stretching vibrations were mathematically deconvoluted into elementary Gaussian profiles with the parameters (the maximum position and full width at half-maximum) determined earlier [14, 15]. The details of this procedure were described in [14].

## 3. Results and discussion
### *3.1. Raman scattering investigation of SiO$_x$N$_y$ films*

Phase structure of the SiO$_x$N$_y$ films with different compositions (see Table 1) was analyzed by Raman spectroscopy. Figure 1(a) shows a number of representative Raman spectra of these films. As can be seen from this figure, the spectra contain two asymmetric bands in the spectral range of 300-800 cm$^{-1}$ with the maxima at ~476 cm$^{-1}$ and ~660 cm$^{-1}$. The first band is associated with Si–Si and Si–O bond vibrations, while the second band corresponds to vibrations of Si–H bonds in an amorphous phase [18-20]. The shape and the frequency position of these bands evidence amorphous structure of our films and presence of silicon-hydrogen bonds, which is typical of the films obtained by PECVD using SiH$_4$ gas [9, 14, 18, 21].

As can be further seen from Figure 1(a), the wide asymmetric band in the frequency range ~ 300-550 cm$^{-1}$ is the most intense one in the measured Raman spectra. The maximum position of this band is always fixed at ~476 cm$^{-1}$, while its half-width Γ decreases with the increase in the relative Si content in the films. We analyzed the shape of this band by decomposing it into Gaussian components (see inset in Figure 1(a)) according to the procedure described in [18, 19, 22, 23]. As a result, two Gaussian components with the parameters $\omega_1$ = 476 cm$^{-1}$ (maximum position) and $\Gamma_1$ = 66 cm$^{-1}$ (half-width), and $\omega_2$ = 420 cm$^{-1}$ and $\Gamma_2$ = 124 cm$^{-1}$ were identified.

As is well known, in Raman spectroscopy, spectral position and shape of a band (width, symmetry, and frequency at maximum) are characteristic parameters that provide important information about the structural state of a studied material. For Si oxide based structures containing amorphous and crystalline clusters, Raman peaks at ~480 cm$^{-1}$ and with a half-width of ~60-80 cm$^{-1}$ corresponding to amorphous silicon TO mode, together with very narrow peaks with the maximum at ~520 cm$^{-1}$ and a half-width of ~5 cm$^{-1}$ corresponding to crystalline Si were observed [20, 22]. Amorphous Si dioxide (a-SiO$_2$) exhibited a broad peak at ~495 cm$^{-1}$ corresponding to the vibrational mode of four-member SiO$_2$ rings [24]. A shift of the spectrum peak from that typical of SiO$_2$ to the one typical of a-Si, its broadening and appearance of a one-side asymmetry were observed for $SiO_x$ films with excess Si content ($x < 2$) at decreasing $x$, indicating heterogeneity of the film structure and change in its phase composition [20, 22]. Raman spectrum of amorphous Si is characterized by broad, unstructured bands in the range of 100-550 cm$^{-1}$ associated with transverse acoustic (TA, ~145 cm$^{-1}$), transverse optical (TO, ~475 cm$^{-1}$), longitudinal optical (LO, ~365 cm$^{-1}$) and longitudinal acoustic (LA, ~305 cm$^{-1}$) phonon vibrations of Si–Si bonds [19, 22, 25]. Therefore, the elementary band at $\omega_1 \approx 476$ cm$^{-1}$ observed in our structures (see inset in Figure 1(a)) may be attributed to the contribution of the TO mode of Si–Si bond vibrations in amorphous silicon. The broader band at $\omega_2 \approx 420$ cm$^{-1}$ may be caused by Raman scattering of light due to Si–O bond vibrations in the amorphous Si oxide matrix [20, 26].

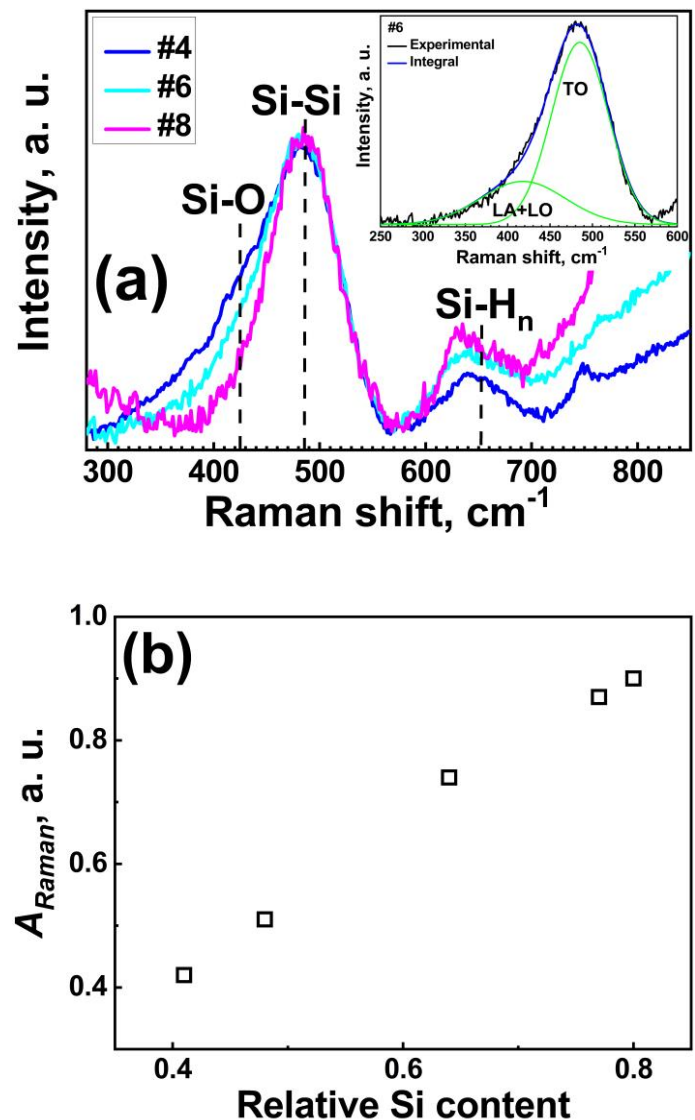

Figure 1. (a) – Raman spectra of $SiO_xN_y$ films obtained at different $N_2O/SiH_4$ gas ratios. Inset – Raman spectrum of the #6 sample deconvoluted into Gaussian components. (b) – Integrated intensity of the $\omega(TO)$ band of the Raman spectra versus Si content in the $SiO_xN_y$ films.

Figure 1(b) shows the dependence of the relative contribution of the integrated intensity of the TO band to the analyzed Raman spectrum in the range of ~300 to 550 cm$^{-1}$, on the relative Si content in the studied SiO$_x$N$_y$ films. As can be seen from this figure, there is a direct correlation between this intensity and the amount of Si in the films. It should be concluded therefore that an increase in the relative Si content in non-stoichiometric Si oxynitride films gives rise to an increase in the amount of amorphous Si in them.

### *3.2. IR spectroscopy study of SiO$_x$N$_y$ film structure*

IR transmittance spectra in the frequency range of 400-4000 cm$^{-1}$ of the SiO$_x$N$_y$ films with different compositions are presented in Figure 2. It can be seen from this figure that the spectra contain transmission bands peaked at ~452, 660, 870, and 1060 cm$^{-1}$ in the low-frequency region and at ~2100 and 3600 cm$^{-1}$ in the high-frequency region. The bands in the range between 900 and 1300 cm$^{-1}$ as well as at ~452 cm$^{-1}$ correspond to Si–O–Si valence and rocking vibrations, respectively [1]. The band peaked at ~870 cm$^{-1}$ is associated with Si–N bonds and the band at ~660 cm$^{-1}$ corresponds to bending vibrations of Si–H bonds [2]. It should be noted at this that the bands in the range of ~600 to 1300 cm$^{-1}$ corresponding to Si–O, Si–N and Si–H bond vibrations can overlap, especially for the films with relatively high Si contents, which greatly complicates analysis of the IR spectra and, hence, the SiO$_x$N$_y$ film structure. The high-frequency bands with the maxima at ~2100 and 3600 cm$^{-1}$ in the measured IR transmittance spectra are caused by valence vibrations of Si–H bonds and Si–OH groups, respectively [1].

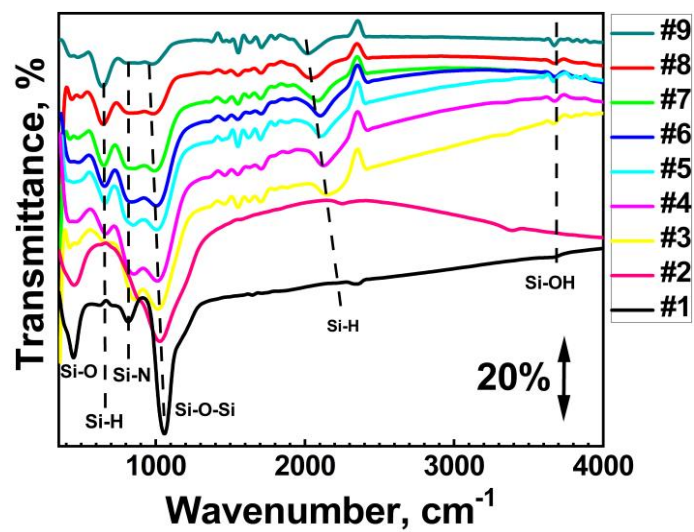

Figure 2. FTIR transmittance spectra of SiO$_x$N$_y$ films with different compositions.

In this work, we focus on a study of IR absorption bands associated with Si–H bond vibrations. Figure 3 shows the high-frequency Si–H related band for a number of the studied

SiO$_x$N$_y$ films. It can be seen from this figure that the shape of this band, its maximum position, intensity and area strongly depend on the SiO$_x$N$_y$ film composition. Namely, increase of the relative Si content in the films leads to the shift of the band peak to lower frequencies, and this shift can be quite large. It should be noted as well that no high-frequency hydrogen-related band is detected in the IR absorbance spectra for the sample with the highest oxygen content and, hence, the lowest Si content (sample #1, see Table 1).

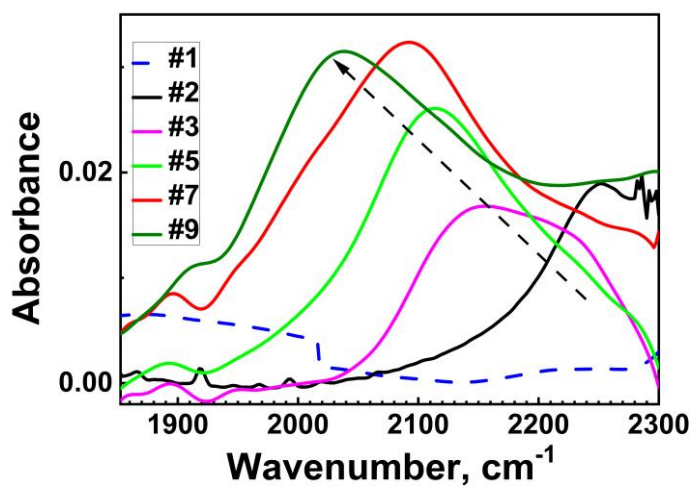

Figure 3. IR absorbance spectra of the SiO$_x$N$_y$ films in the range of Si–H stretching vibrations. The arrow shows the direction of the shift of the maximum position of the integrated band with the increase in the relative Si content in the films.

By integrating the area under the IR spectra in the range of ~1900-2400 cm$^{-1}$, the amount of hydrogen bound in the studied films can be calculated [14, 15]. The calculated hydrogen concentration as a function of the relative Si content in the films is presented in Figure 4. It can be seen from this figure that the hydrogen concentration is of the order of ~10$^{21}$-10$^{22}$ cm$^{-3}$, which coincides by the order of magnitude with the known literature data (see e.g. [7]). This concentration increases with the relative amount of Si in the films, which looks quite natural, since growth of the latter is associated with an increase in the SiH$_4$ concentration in the gas mixture during the film deposition. For the samples with the highest Si content, a tendency to saturation is noticeable.

Composition of the IR spectral band in the range of ~1900-2400 cm$^{-1}$ allows one to draw fundamental conclusions about the structural arrangement of hydrogen atoms in the Si oxynitride network. It is well-known that the considered wide band is a superposition of elementary bands corresponding to stretching modes of H–Si(Si$_{3-n}$O$_n$) structural units, where $n$ = 0…3 [14, 15]. Hence, we deconvoluted this band into elementary Gaussian components for different Si oxynitride film stoichiometries taking into account both the nature of the elementary components [15] as well as our previous experience gained during the study of the kinetics of hydrogen effusion from Si-rich

Si oxynitride films induced by heat treatments [14]. The deconvolution results for the films depicted in Figure 3 are presented in Figure 5. As can be seen from this figure, three elementary components, namely H1, H2 and H3 with the characteristics presented in Table 2, were separated. As can be further seen from Figure 5, the spectrum of the sample #2 having low Si content (see Table 1) is described by the only component H1. Increase in the relative Si content leads to redistribution of the main contribution from H1 to H2 (for the sample #3) and H3 (for the samples #7 and #9) with a tendency of gradual disappearance of the previous bands. Figure 6 provides a more detailed information about the dependence of the relative contributions of the bands H1, H2 and H3 to the FTIR spectra of the $SiO_xN_y$ films on the relative Si content in the films for all the samples presented in Table 1.

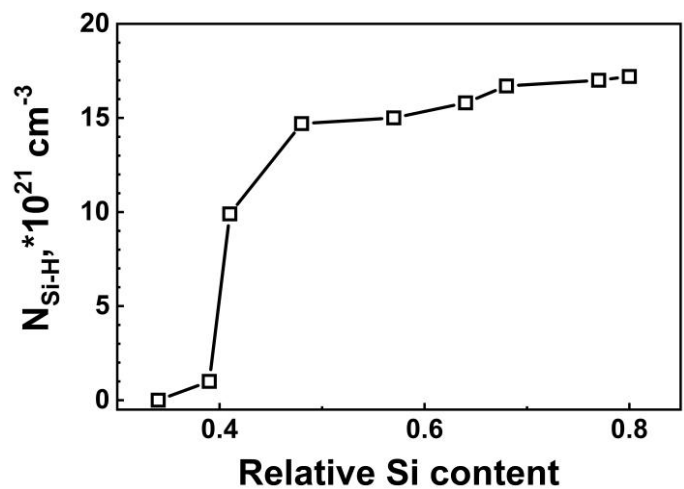

Figure 4. Hydrogen concentration versus relative Si content in the $SiO_xN_y$ films.

It is known [4, 15] that the quantity and type of the neighboring atoms to a hydrogen atom correlate with the Si–H bond length, and, hence, influence on the Si–H stretching vibrations frequency. For each of the separated elementary bands H1 to H3, the Si–H bond length was determined using the following expression [4]:

$$\nu_{Si-H}(d_{Si-H})^3 = 7074 \text{ cm}^2 \tag{1}$$

where $\nu_{Si-H}$ is the band peak position and $d_{Si-H}$ is the Si–H bond length, respectively. The obtained bond lengths together with the respective $H-Si(Si_{3-n}O_n)$ complexes identified according to [15] are presented in Table 2. It should be noted that the Si–H bond length for the bands H1 (0.146 nm) and H3 (0.151 nm) determined by the expression (1) practically do not differ from the respective values for Si dioxide (0.144 nm) and amorphous Si (0.152 nm) as indicated in [4]. We may conclude therefore that the results of the analysis of the high-frequency parts of the IR spectra agree well with existence of amorphous Si phase within the Si oxynitride structure at the relative Si content above about 0.4 (samples #3 to #9, see Figure 6).

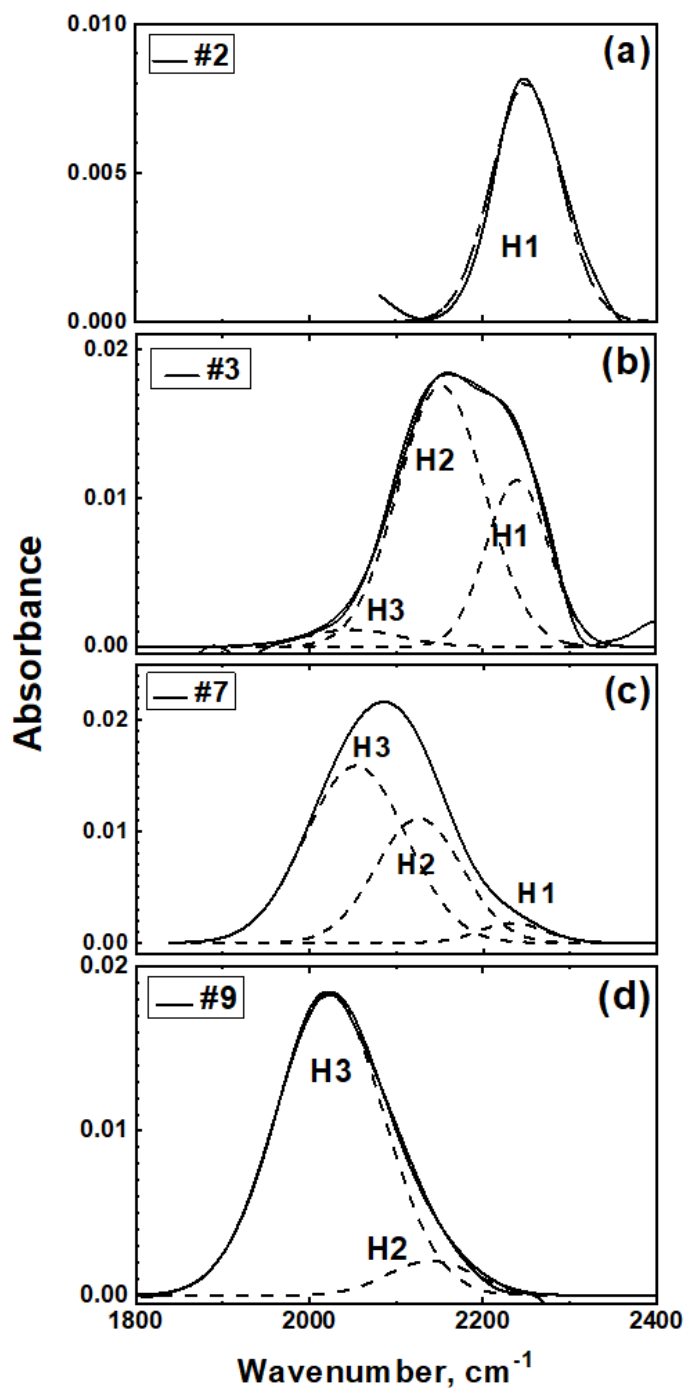

Figure 5. Deconvolution of the IR absorption band corresponding to Si–H stretching vibrations on elementary Gaussian components corresponding to contributions from H–Si(O$_3$) (H1), H–Si(Si$_2$O) (H2) and H–Si(Si$_3$) (H3) complexes.

We have mentioned above that another Si–H related band in the measured IR absorbance spectra corresponds to the peak position of about 660 cm$^{-1}$. This band is related to Si–H bending vibrations [9, 14-16]. Analysis of the integrated intensity of this band would provide additional

**Table 2.** Characteristics of the Gaussian components of the Si–H related IR absorption band

| Band | Maximum position, cm$^{-1}$ | Width, cm$^{-1}$ | Oscillation type | Complex | $d_{Si-H}$, nm |
|---|---|---|---|---|---|
| H1 | 2252±2 | 75±3 | stretching | H–Si(O$_3$) | 0.146 |
| H2 | 2150±4 | 100±2 | stretching | H–Si(Si$_2$O) | 0.149 |
| H3 | 2050±4 | 85±4 | stretching | H–Si(Si$_3$) | 0.151 |

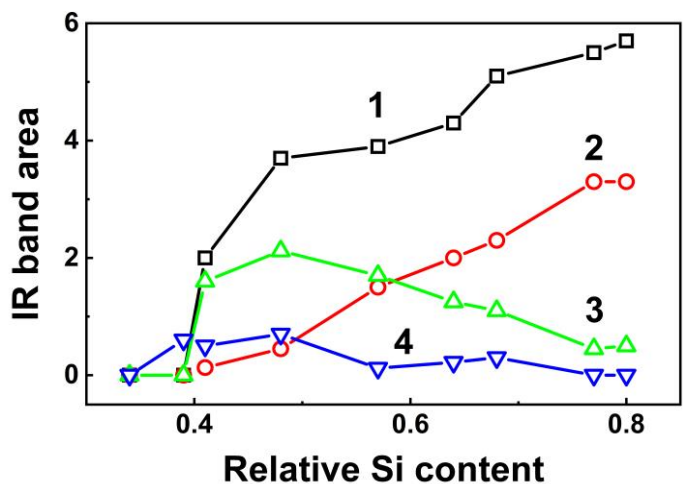

Figure 6. Integrated intensity of IR absorbance bands corresponding to Si–H vibrations versus Si content in the SiO$_x$N$_y$ films. The maximum position of the IR bands: 1 – ~660 cm$^{-1}$, 2 – ~2050 cm$^{-1}$, 3 – ~2150 cm$^{-1}$, and 4 – ~2250 cm$^{-1}$.

information about structural arrangement of hydrogen in the Si oxynitride films as well as highlight the Si oxynitride phase composition. However, analysis of this band is generally complicated as the band may overlap with the bands corresponding to Si–O and Si–N bond vibrations also present in the range of ~600 to 1300 cm$^{-1}$. In our previous publication [13], we determined the parameters of the elementary Gaussian components of the IR absorbance spectra of Si-rich Si oxynitride films in the mentioned frequency range thus enabling separation of the Si–O, Si–N and Si–H related bands. Figure 7(a) shows the IR absorbance spectra of a number of SiO$_x$N$_y$ samples in the range of 500 to 800 cm$^{-1}$ with separated absorption band near ~660 cm$^{-1}$ corresponding to Si–H bond vibrations. This band is described by a single Gaussian with the mean half-width of ~82 ± 2 cm$^{-1}$. It can be seen from Figure 7(a) that the intensity and maximum position of this band depend on the N$_2$O/SiH$_4$ flow ratio during Si oxynitride film growth and, hence, the relative Si content in the films. The dependence of the integrated intensity of this band on the relative Si content in the studied films is also presented in Figure 6. As can be seen from Figures 6 and 7(a), the mentioned band is absent in the IR spectra of the films with the smallest Si contents (samples #1 and #2) and increases its integrated intensity with subsequent growth of the relative Si amount (samples #3 to #9). The position of the band maximum shifts at this from ~665 cm$^{-1}$ for the sample #3 (relative Si content of

0.41) to ~635 cm$^{-1}$ for the sample #9 (relative Si content of 0.8), as shown by the data provided in Figure 7(b). These findings agree well with the results of [4, 18], where the low-frequency IR band corresponding to Si–H bending vibrations in PECVD grown Si oxynitride films had no definite peak position. We conclude therefore that there is a clear correlation between the position of the peak and the integrated intensity of the IR absorption band at ~660 cm$^{-1}$ corresponding to Si–H bond bending vibrations on the one hand, and the Si content in the SiO$_x$N$_y$ samples on the other hand.

We have established above that the IR absorption band at ~2050 cm$^{-1}$ corresponds to the Si–H bond stretching vibrations in amorphous Si phase. It can be seen from Figure 6 that the integrated intensity of this band has a direct correlation with that of the band at ~660 cm$^{-1}$. Moreover, as can be seen from Figure 8, the integrated intensity of the latter band is directly proportional to the integrated intensity of the TO mode of Si–Si bond vibrations in the measured Raman spectra. Therefore, the IR absorption band peaked at ~640-660 cm$^{-1}$ may be confidently attributed to the presence of amorphous Si phase in the non-stoichiometric Si oxynitride films. This conclusion is supported by the published results by other researchers. It is reported in particular that the IR absorption band with the peak position around 640 cm$^{-1}$ is typical for a-Si layers [27] and is considered to origin from Si–H bond vibrations in H–Si(Si$_3$) complexes [16, 17]. In Si oxynitride films, this band was reliably detected only in the case of high relative Si contents. It was not detected in the spectra of the samples deposited by PECVD technique, which had a rather high oxygen concentration and a composition much closer to SiO$_2$ than to amorphous Si [15]. On the other hand, a band peaked at ~640 cm$^{-1}$ corresponding to Si–H bending vibrations was present in the IR spectra of as-deposited Si-rich Si oxide films [13]. Moreover, an absorption band with the maximum near 670 cm$^{-1}$ clearly manifested itself in the IR spectra of the SiO$_x$N$_y$ samples with high Si fraction [4]. In both latter cases, presence of a-Si inclusions in the studied films was established independently.

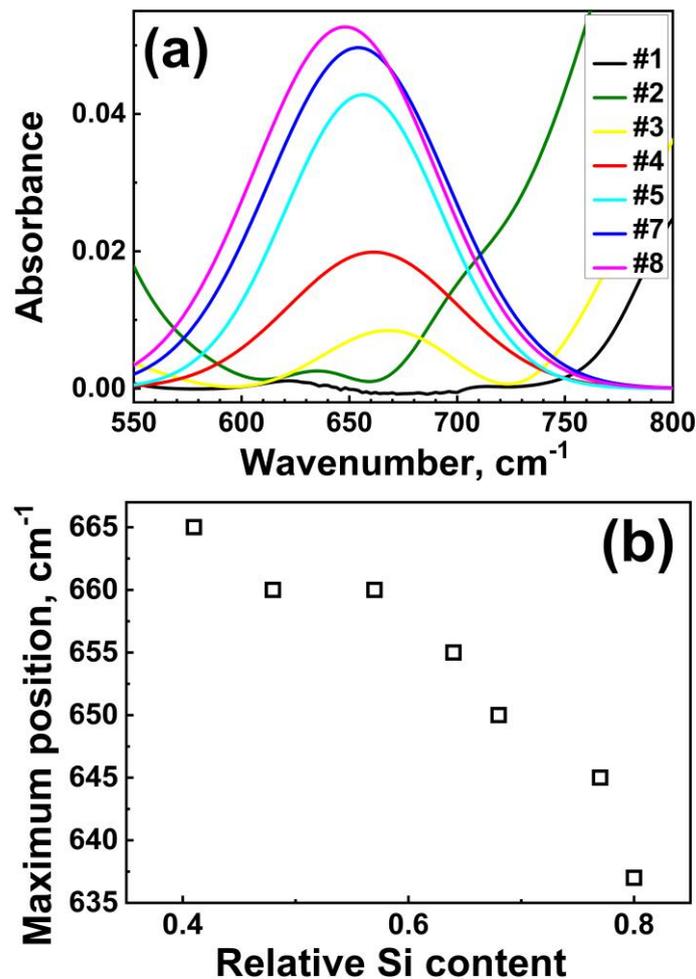

Figure 7. (a) – FTIR absorbance spectra of the SiO$_x$N$_y$ films corresponding to Si–H bending vibrations and (b) – dependence of the maximum position of the band at ~660 cm$^{-1}$ on the relative Si content in the SiO$_x$N$_y$ films.

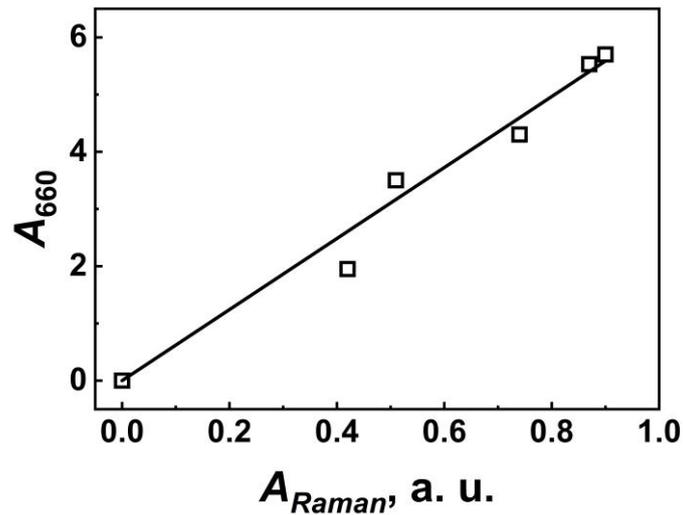

Figure 8. Integrated intensity of the IR absorption band at ~660 cm$^{-1}$ versus integrated intensity of the ω(TO) band of the Raman spectra of the SiO$_x$N$_y$ films.

### 3.3. Comparative analysis of IR and Raman scattering results

Comparative analysis of the evolution of the Si–H related IR bands with the SiO$_x$N$_y$ composition along with the respective Raman scattering results allows one to get a deeper insight into the peculiarities of the formation of phase composition of Si oxynitride films grown by plasma-enhanced chemical vapor deposition.

As demonstrated by the data from [4, 9, 15, 21] as well as our Raman and IR spectroscopy investigations, formation of amorphous Si clusters in an oxynitride matrix is possible in Si oxynitride films deposited by PECVD at different N$_2$O/SiH$_4$ gas flow ratios. In particular, presence of a-Si phase is detected by Raman measurements (observation of the TO mode of Si–Si bond vibrations at $\omega_1 \sim 476$ cm$^{-1}$, see Figure 1(a)). At this, the amount of amorphous Si grows with the increase in the relative Si content in the films.

Using the approach to deconvolution of IR absorbance spectra of Si oxynitride films presented in [13], a low-frequency IR band corresponding to Si–H bending vibrations (~640-660 cm$^{-1}$) is reliably separated. As was already mentioned above, the integrated intensity of this band correlates both with the intensity of the IR absorption band at ~2050 cm$^{-1}$ identified to origin from the Si–H stretching vibrations in amorphous Si (see Figure 6), as well as with the integrated intensity of the TO mode in the Raman scattering spectra (see Figure 8), which made it possible to reliably attribute this band to the presence of a-Si phase. As can be further seen from Figure 6, no a-Si related signal is detected in the IR spectra of the samples with small relative Si concentrations (below about 0.4). This points to a certain threshold value of excess Si content, at which amorphous Si phase begins to form. This is further demonstrated by Figure 9, where the dependence of the integrated intensity of the IR absorption band at ~660 cm$^{-1}$ on the excess Si content in the investigated Si oxynitride films calculated as $1 - x/2 - 3y/4$ is shown. As can be seen from this figure, the threshold excess Si value is about 0.2, and the amount of the amorphous Si in the Si oxynitride films, which is proportional to the integrated intensity of the IR band at ~660 cm$^{-1}$, linearly grows with the excess Si concentration above this value.

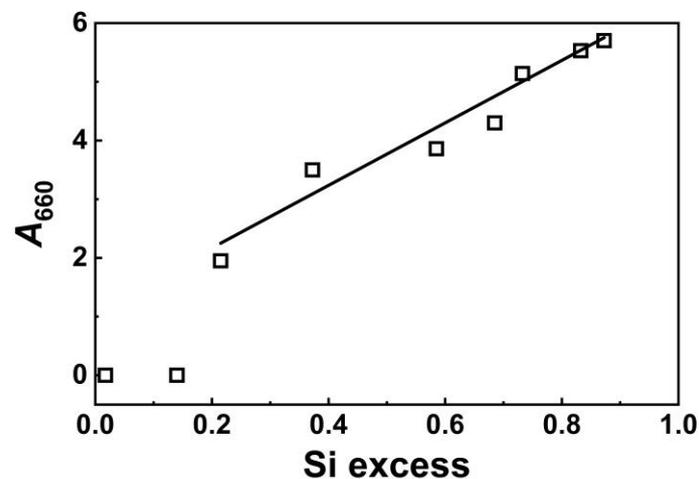

Figure 9. Integrated intensity of the IR absorption band at ~660 cm$^{-1}$ versus excess relative Si content in the SiO$_x$N$_y$ films.

The properties of the IR absorption band at ~660 cm$^{-1}$ allow one to consider IR spectroscopy as an efficient method for tracking Si agglomeration into amorphous Si phase in an oxynitride matrix and formation of a particular phase composition of PECVD grown films. According to this method, IR transmittance spectra of SiO$_x$N$_y$ films in the low-frequency region should be only measured. Using the deconvolution procedure and the characteristics of the elementary IR bands corresponding to Si–O, Si–N and Si–H (~660 cm$^{-1}$) bond vibrations [13], phase composition of the studied films, including the presence or absence of amorphous Si inclusions in the films, may be inferred. We believe that this method is one of the simplest and most non-destructive ones for monitoring Si agglomeration in SiO$_x$N$_y$ films produced by PECVD technology and determining their phase composition and structure.

The results obtained in this study extend our understanding of the peculiarities of formation of phase compositions of PECVD grown Si-rich Si oxynitride films. Recently, a thermodynamic theory describing phase composition of SiO$_x$N$_y$ films with different stoichiometries grown at different temperatures was proposed [28], which, however, considered only a random bonding model distribution of Si, O and N atoms. Identification of the possibility of presence of amorphous Si clusters in such films should be used for updating the theoretical model, in particular, to improve its potential for tailoring the phase composition properties of Si oxynitride films for practical applications.

**4. Conclusions**

In this work, peculiarities of formation of amorphous Si phase in Si oxynitride films grown by plasma-enhanced chemical vapor deposition are studied by combined Raman scattering and IR spectroscopy investigations. The results of the Raman spectroscopy demonstrate that at a certain ratio of precursor gases in the growth chamber during SiO$_x$N$_y$ film deposition, excess Si content gives rise to formation of amorphous Si phase within the films. This is evidenced by the appearance of the peak at ~476 cm$^{-1}$ corresponding to the transverse optical phonon mode characteristic of a-Si. Large half-width of this band ($\Gamma$ = 66 cm$^{-1}$) indicates the absence of long-range order in the film structure, which is typical for amorphous materials. The results of our Raman spectroscopy investigations are in good agreement with the XPS, XRD, and PL results reported in [4, 6, 9, 21] for similar Si oxynitride films prepared by PECVD technology.

Using the method of deconvolution of IR absorbance spectra in the wavelength range of ~600 to 1300 cm$^{-1}$ into elementary components, proposed in [13], the elementary band at ~660 cm$^{-1}$ corresponding to Si–H bending vibrations was separated. The comparative analysis of the IR and Raman spectroscopy results enabled identification of this band as related to the presence of amorphous Si inclusions in the studied films. It was demonstrated that formation of amorphous Si clusters in Si oxynitride films takes place at the relative excess Si concentrations exceeding a threshold value of about 0.2. After the threshold, the amount of the a-Si phase linearly grows with the excess Si concentration. This result makes it possible to consider IR spectroscopy with analysis of the low-wavelength (~600 to 1300 cm$^{-1}$) part of spectra as an efficient method of identifying phase composition of Si-rich Si oxynitride films.

The results of this study enhance our understanding of the peculiarities of the formation of phase compositions of PECVD grown Si oxynitride films and may be used for developing methods of creating the films with required phase compositions for practical applications.

**Acknowledgement**

This work has been supported by the budget project no. III-4-21 "Electrical, optical and photoelectrical characteristics of the systems with nanostructured surfaces and nanocrystals; physical mechanisms of emission conversion in modern optoelectronic structures for developing novel optoelectronic, sensoric, energy saving illumination, and information recording and storage devices" of the National Academy of Sciences of Ukraine.